\documentclass{elsart}
\usepackage{epsfig}
\begin{document}
\begin{frontmatter}
\title{Itinerant Antiferromagnetism in $\mathrm{FeGe_2}$}

\author{T.E. Mason, C.P. Adams, S.A.M. Mentink, E. Fawcett}
\address{Department of Physics, University of Toronto, Toronto, ON M5S 1A7, 
Canada}
\author{A.Z. Menshikov}
\address{Institute for Metal Physics, Ekaterinburg, Russia}
\author{C.D. Frost, J.B. Forsyth, T.G. Perring}
\address{ISIS Facility, Rutherford Appleton Laboratory, Chilton,
Didcot, OX11 0QX, United Kingdom}
\author{T.M. Holden}
\address{AECL, Chalk River Laboratories, Chalk River, ON K0J 1J0, Canada}

\begin{abstract}
$\mathrm{FeGe_2}$, and lightly doped compounds based on it, have a Fermi
surface driven instability which drive them into an incommensurate spin
density wave state.  Studies of the temperature and magnetic field dependence
of the resistivity have been used to determine the magnetic phase diagram
of the pure material
which displays an incommensurate phase at high temperatures and a
commensurate structure below 263 K in zero field.  Application of a magnetic 
field in the tetragonal basal plane decreases the range of temperatures
over which the incommensurate phase is stable.  We have used inelastic
neutron scattering to measure the spin dynamics of $\mathrm{FeGe_2}$.
Despite the relatively isotropic transport the magnetic dynamics is 
quasi-one dimensional in nature.  Measurements carried out on HET at ISIS
have been used to map out the spin wave dispersion along the c-axis up
the 400 meV, more than an order of magnitude higher than the zone boundary
magnon for wavevectors in the basal plane.
\end{abstract}
\end{frontmatter}

\newpage
Studies of the spin dynamics of the high T$_{c}$ 
superconductor,
${\rm La_{2-x}Sr_{x}CuO_{4}}$, have shown that this material has 
incommensurate magnetic
fluctuations characteristic of a metal close to a spin density wave
instability\cite{mason94}. 
This has led to a renewed interest in systems which exhibit similar 
instabilities since there are many unanswered questions regarding the
transport and magnetic properties of spin density wave systems.  The 
prototypical SDW system is the itinerant antiferromagnet Cr and dilute
alloys of Cr with other transition metals\cite{fawcett94}.  SDW 
ordering also occurs in Mott-Hubbard systems where strong correlations
lead to a break down of conventional band theory and, as in the high 
${\rm T_{c}}$ cuprates, a metal insulator transition\cite{bao93}.
$\mathrm{FeGe_2}$ and lightly doped compounds based on it have a 
similar (Fermi surface) spin density wave instability which drives the 
system to order and, in its
undoped form, exhibits both commensurate and incommensurate magnetic
ordering as temperature is varied.  
We are studying the transport and spin dynamics of this system in 
order to obtain a better microscopic understanding of 
how this instability is manifested in the spin fluctuations.

The properties of $\mathrm{FeGe_{2}}$, a tetragonal material with the
$\mathrm{CuAl_2}$ structure, have been studied 
since 1943\cite{walbaum43} and many of the details of the
crystal and magnetic structure have been determined
\cite{corliss85,dorofeyev87,menshikov88,forsyth64}.  $\mathrm{FeGe_{2}}$ is
characterised by two magnetic phase transitions, the first of which is the
N\'{e}el transition from a paramagnetic phase to an incommensurate spin
density wave state at $\mathrm{T_N}$ = 289 K 
in zero field.  The second one is a
commensurate-incommensurate (C-IC) transition into
a collinear antiferromagnetic structure.  This 
transition occurs at
$\mathrm{T_c}$ = 263 K in zero field.  
The propagation vector $\vec {Q}$ of the SDW is parallel to [100]
and varies
from 1 to roughly 1.05 $a^{*}$ between the two transitions
which are seen in
resistivity\cite{krentsis70,krentsis73}, ultrasonic 
attenuation\cite{pluznikov82},
heat capacity\cite{corliss85,mikhelson71}, and AC
susceptibility\cite{tarasenko89} studies.  The spins lie in the basal plane
and are ferromagnetically aligned along the c-axis.  
Although this compound has been well characterised in zero field there 
has been very little work done in magnetic fields and not much 
is known about the details of the magnetic phase diagram.
The magnon and phonon dispersion were measured (primarily at low
temperatures) and revealed a significant anisotropy of the magnetic
interactions with the spin wave velocity along the c-axis much 
higher than that observed for modes propagating in the basal plane
\cite{holden96}.  For those thermal neutron triple-axis measurements it was 
not possible to fully characterise the spin wave dispersion along c due to 
the unsuitability of that technique for energy transfers beyond about 50 meV.

\begin{figure}
\begin{center}
\epsfig{file=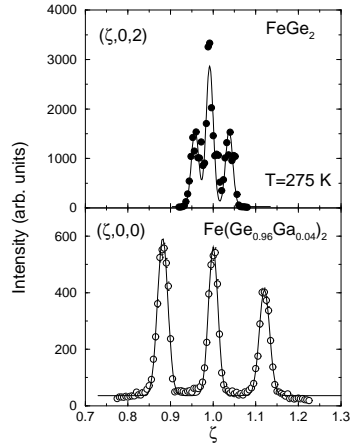,height=7cm}
\end{center}
\caption[qscan]{Scans through the incommensurate magnetic Bragg peaks for
pure (upper panel) and Ga doped (lower panel) $\mathrm{FeGe_{2}}$ at 275 K
obtained on the E3 and C5 triple axis spectrometers at Chalk River.  The peak
in the centre arises from the two peaks displaced from the commensurate 
position along the b-axis, seen due to the coarse vertical collimation 
employed in these measurements.}
\label{qscan}
\end{figure}
Incommensurate ordering of the type seen in $\mathrm{FeGe_{2}}$ gives rise to
four magnetic satellites around (100) and related positions occurring
at $(1\pm q,0,0)$ and $(1,\pm q,0)$.  
Elastic neutron scattering can be used to determine the wavevector of
this magnetic ordering, a scan through the incommensurate peaks at 275 K is 
shown in the upper
panel of Figure 1 compared to that of $\mathrm{Fe(Ge_{0.96}Ga_{0.04})_2}$
at the same temperature.  
Doping has the effect of increasing the magnitude 
of the incommensurate wavevector (similar to what is seen in lightly
doped alloys of Cr\cite{fawcett94}).  In the 4 at\% Ga doped sample the
ordering wavevector varies between $q=0.12$ just 
below $\mathrm{T_N=285 K}$ 
and $q=0.1$ below a first order lock-in transition at 
$\mathrm{T_c=182 K}$.
In the pure material the transition to the low temperature commensurate
state (with $\mathbf{q}=0$) is continuous.  
\begin{figure}
\begin{center}
\epsfig{file=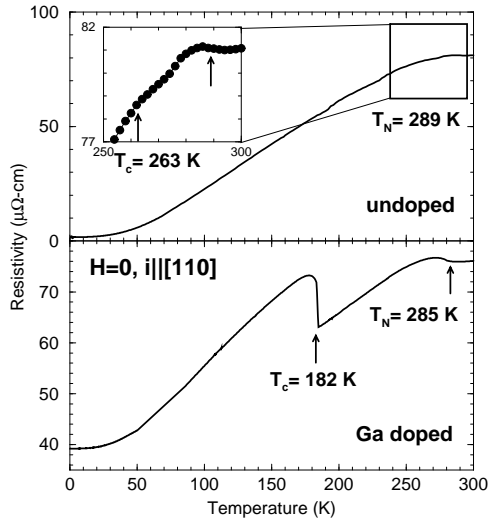,height=7cm,bbllx=100,bblly=100,bburx=452,bbury=550}
\end{center}
\caption[resist]{Temperature dependence of the resistivity for pure and 
Ga doped $\mathrm{FeGe_{2}}$ (measured on pieces of the same samples
used for Figure 1).}
\label{resist}
\end{figure}
The zero field resistivity for
these two compounds is shown in Figure 2.  In the doped sample (bottom
panel) the two phase transitions are clearly visible and are denoted by arrows.
In the pure material the anomalies are more subtle but the transitions are
seen in the inset.  The transition temperatures deduced from the resistive
anomalies agree well with those determined by neutron scattering.  
\begin{figure}
\begin{center}
\epsfig{file=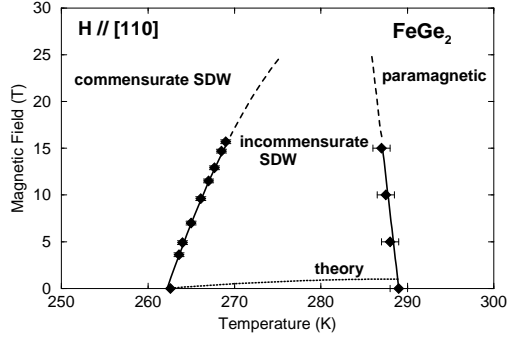,height=5cm,bbllx=100,bblly=220,bburx=452,bbury=570}
\end{center}
\caption[pdiag]{Magnetic phase diagram for $\mathrm{FeGe_{2}}$ for a
magnetic field applied along the [110] direction determined by resistivity
measurements and (for H=0) neutron diffraction.  The line denoted theory
corresponds to the approximate phase boundary expected based on the
Ginzburg-Landau theory of Reference \cite{tarasenko89}}
\label{pdiag}
\end{figure}
We have
used similar measurements, together with constant temperature, field scans
which clearly show the lower phase transition, with a magnetic field applied 
in the basal plane to determine the
magnetic phase diagram for $\mathrm{FeGe_{2}}$, shown in Figure 3, the 
behaviour is similar for a field applied along [100].\cite{adams96}

\begin{figure}
\begin{center}
\epsfig{file=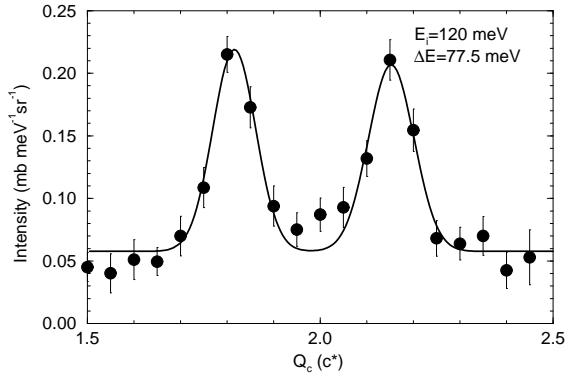,height=5cm,bbllx=100,bblly=180,bburx=452,bbury=505}
\end{center}
\caption[conste]{Constant energy cut at 77.5 meV 
through the two counter-propagating
spin wave modes obtained on HET for $\mathrm{FeGe_{2}}$ at 11 K using
an incident neutron energy of 120 meV.}
\label{conste}
\end{figure}
In order to fully characterise the ground state of $\mathrm{FeGe_{2}}$ we
have measured the spin dynamics at low temperatures (11 K) using the
direct geometry time-of-flight spectrometer HET at ISIS.  This instrument
allows access to much higher incident neutron energies than the
thermal triple-axis measurements of Holden et. al.\cite{holden96} making
it possible to fully map out the c-axis dispersion.  Figure 4 shows a 
constant energy cut through the time-of-flight trajectories of the 2.5 m
detector bank on HET for an energy transfer of 77.5 meV.  This was obtained
for an integrated current of 2200 $\mu$A-hrs using an incident energy of
120 meV.  The two spin wave modes emanating from $\mathrm{Q_c}=2$ are
clearly visible, for energies higher than 25 meV $\mathrm{FeGe_{2}}$ behaves
as a quasi-one-dimensional ferromagnet so it is permissible to ignore 
the variation of $\mathrm{Q_a}$ with energy transfer, a fact confirmed by 
more detailed analysis of these results\cite{adams97}.
\begin{figure}
\begin{center}
\epsfig{file=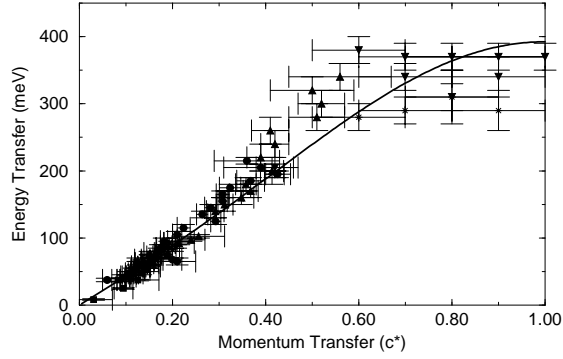,height=5cm,bbllx=100,bblly=150,bburx=452,bbury=475}
\end{center}
\caption[cdisp]{Spin wave dispersion relation obtained from the triple
axis data (lowest energies) and incident energies of 80, 120, 300, 500,
and 700 meV on HET.  The line is a fit to a model for a planar 
one-dimensional ferromagnet with a zone boundary magnon energy of 400 meV.}
\label{cdisp}
\end{figure}
By carrying out measurements with incident energies of 80, 120, 300, 500,
and 700 meV we have fully mapped out the c-axis dispersion relation which
is shown in Figure 5.  The line is the result of a fit to
a planar one-dimensional ferromagnet model in appropriate in the zero
temperature classical limit.  The zone boundary magnon energy is 400 meV, more
an order of magnitude larger than the 25 meV found for the 
a-axis\cite{holden96}.  This substantial anisotropy reflects the structure of
$\mathrm{FeGe_{2}}$, the Fe-Fe distance in the c-axis chains of Fe spins is
similar to that in pure Fe, whereas in the ab-plane the distance between
the Fe ions is 4.2 \AA.  Along the c-axis the interaction is ferromagnetic and 
mainly due to d-orbital overlap while in the basal plane the weaker 
antiferromagnetic coupling has its origins in the conduction electrons.

$\mathrm{FeGe_{2}}$ is a unique example of quasi-one-dimensional ferromagnetism
in an itinerant electron system.  Studies of the magnetic dynamics at higher
temperatures\cite{adams97} reveal the presence of well resolved magnon-like
excitations even in the paramagnetic state.  The much weaker basal plane
coupling drives the three dimensional ordering, the doping dependence of
the ordering wavevector suggests that this is Fermi surface driven, hole
doping by substituting Ga for Ge increases the incommensurate wavevector
similar to the effect of increased Sr doping in $\mathrm{La_{2-x}Sr_xCuO_4}$
\cite{mason94} or V doping in $\mathrm{Cr_xV_{1-x}}$\cite{fawcett94}.
which are quasi-two-dimensional and three-dimensional realizations of systems
which exhibit Fermi surface driven spin density wave instability. The
$\mathrm{Fe(Ge_{1-x}Ga_x)_{2}}$ alloy system represents another such system
which is predominantly one-dimensional magnetically although the 
relatively small anisotropy in the resistivity (a factor of two with higher
resistivity along c) indicates the charge carriers are not
restricted to the c-axis.

We would like to thank the technical staff of ISIS and Chalk River Laboratories
for expert assistance with the neutron scattering measurements described in
this paper.  This work was financially supported by NSERC and the CIAR.

\end{document}